\documentclass[review]{elsarticle}

\begin{document}

\begin{frontmatter}

\title{Novel multi-layer plastic-scintillator-based solid active proton target for inverse-kinematics experiments \tnoteref{mytitlenote}}
\tnotetext[mytitlenote]{The Super-FRS Collaboration}

\author[a,b]{D.~T.~Tran}
\author[c]{S.~Terashima}
\author[a]{H.~J.~Ong\corref{mycorrespondingauthor}}
\cortext[mycorrespondingauthor]{Corresponding author}
\ead{onghjin@rcnp.osaka-u.ac.jp}
\author[a]{K.~Hirakawa}
\author[d]{Y.~Matsuda}
\author[a]{N.~Aoi}
\author[e,f]{M.~N.~Harakeh}
\author[d]{M.~Itoh}
\author[g]{T.~Kawabata}
\author[a]{A.~Kohda}
\author[h]{S.~Y.~Matsumoto}
\author[i]{T.~Nishi}
\author[d]{J.~Okamoto}
\author[a,c]{I.~Tanihata}

\address[a]{Research Center for Nuclear Physics (RCNP), Osaka University, Ibaraki, Osaka 567-0047, Japan}
\address[b]{Institute of Physics, Vietnam Academy of Science and Technology, Hanoi 10000, Vietnam}
\address[c]{School of Physics and Nuclear Energy Engineering, Beihang University, Beijing 100191, China}
\address[d]{Cyclotron and Radioisotope Center (CYRIC), Tohoku University, Sendai, Miyagi 980-8578, Japan}
\address[e]{KVI Center for Advanced Radiation Technology, University of Groningen, 9747 AA Groningen, The Netherlands}
\address[f]{GSI Helmholtzzentrum fur Schwerionenforschung GmbH, Planckstrasse 1, 64291 Darmstadt, Germany}
\address[g]{Department of Physics, Osaka University, Osaka 560-0043, Japan}
\address[h]{Department of Physics, Kyoto University, Kyoto 606-8502, Japan}
\address[i]{RIKEN Nishina Center, Saitama 351-0198, Japan}


\begin{abstract}
  We have constructed and tested a novel plastic-scintillator-based solid-state
  active proton target for use in nuclear spectroscopic studies with nuclear
  reactions induced by an ion beam in inverse kinematics. The active target
  system, named Stack Structure Solid organic Scintillator Active Target
  (S$^4$AT), consists of five layers of plastic scintillators, each with a
  1-mm thickness. To determine the reaction point in the thickness direction,
  we exploit the difference in the energy losses due to the beam particle and
  the charged reaction product(s) in the scintillator material. S$^4$AT offers
  the prospect of a relatively thick target while maintaining a good energy
  resolution. By considering the relative energy loss between different layers,
  the energy loss due to unreacted beam particles can be eliminated. Such
  procedure, made possible by the multi-layer structure, is essential to
  eliminate the effect of unreacted accompanying beam particles, thus enabling
  its operation at a moderate beam intensity of up to a few Mcps. We evaluated
  the performance of S$^4$AT by measuring the elastic proton-proton scattering
  using a 70-MeV proton beam at Cyclotron and Radioisotope Center (CYRIC),
  Tohoku University.
\end{abstract}


\end{frontmatter}


\section{Introduction}
Nuclei beyond the beta-stability line, commonly known as radioactive isotopes
or unstable nuclei, are currently at the forefront of nuclear physics research.
Over the last few decades, extensive experiments using radioactive-isotope (RI)
beams have revealed various exotic and fascinating features of nuclear
structure and dynamics hitherto unknown in nuclei on or near the
$\beta$-stability line. Of particular interest are the formation of neutron-halo
structure~\cite{Tanihata85} and evolution of nuclear magic
numbers~\cite{Sorlin08,Nakamura17,Tran18} in neutron-rich nuclei. While the
exact mechanisms for both phenomena have remained unclear, it seems that the
tensor interactions may play an important role~\cite{Otsuka05,Myo07}. The
observations of possible effect of the tensor interactions in $^{16}$O via
(p,d)~\cite{Ong13} and (p,dp)~\cite{Terashima18} reactions on $^{16}$O at
high-momentum transfer have opened up new possibilities to understand the
effect of tensor interactions on the structure of atomic nuclei.


The unstable nuclei, with long isotopic chains, provide the ideal platforms for
systematic studies using nuclear reactions. The increasing availability of a
wide range of highly- or moderately-intense RI beam species, and developments
of more complex and sophisticated detector systems such as multi-strip silicon
arrays~\cite{Pollacco05,Wallace07} and active gaseous target detector
systems~\cite{Hashimoto06,Demonchy07,Suzuki12,Ota15,Furuno18} have enhanced
the prospect for spectroscopic studies of unstable nuclei with nuclear
reactions. One of the most versatile experimental methods via nuclear reactions
is the missing-mass spectroscopy. In this method, the momentum/momenta of the
reaction product(s), usually light ion(s), is/are measured. The momentum
information obtained is then used to reconstruct the excitation energy of the
residual nucleus right after the reaction, and to determine physical
observables such as
differential cross sections. To ensure sufficient separation for the ground and
different excited states in the residual nucleus, it is important to attain
good momentum resolution. For experiments with stable beams and stable target
nuclei, a good momentum resolution is usually fulfilled by using a thin target.
The same prescription, however, is seldom applicable to experiments with RI
beams due to their limited beam intensities. Hence, to increase luminosity and
thereby improve the experimental feasibility, it is essential to increase the
target thickness whenever possible. The use of a thicker target, however,
may result in the degradation of the momentum resolution due to the uncertainty
in the depth of reaction vertex.

Here, we report on the development of a novel multi-layer solid-state active
proton target, named Stack Structure Solid organic Scintillator Active Target
(S$^4$AT). The device consists of ordinary commercial plastic scintillators, and
is therefore highly affordable and flexible; such advantages have already been
demonstrated by the recently developed S$^4$ neutron detector~\cite{Yu17}.
S$^4$AT offers a versatile alternative proton (or deuteron) target for
spectroscopic studies with radioactive isotope (RI) beams using nuclear
reactions, especially reactions that result in more than one charged particle
being ejected, in inverse kinematics. Because the total energy loss of charged
particles in the scintillator material differs significantly before and after
such reactions, one can determine the reaction point in the thickness direction
by measuring the energy-loss distribution across different layers of the active
target. This reaction-depth determination capability allows the use of a
relatively thick target while maintaining a relatively good energy resolution.
The multi-layer structure is of importance because it provides a means to
subtract the energy loss due to unreacted beam particles through consideration
of the energy-loss difference between adjacent layers. Such feature is
essential to eliminate the effect of accompanying unreacted beam (pile-up)
particles. These depth-sensitive and pile-up elimination capabilities, together
with the short decay time of the plastic scintillator, enable its operation at a
moderate beam intensity of up to a few Mcps. To assess the performance of
S$^4$AT, we have conducted a test experiment using elastic proton scattering
with a 70-MeV proton beam at Cyclotron and Radioisotope Center (CYRIC), Tohoku
University.

The article is organized as follows. In Section~\ref{sec:principle}, we
describe the basic principle for vertex determination and the prototype system.
The test experiment is presented in Section~\ref{sec:exp}. The data analysis and
results are presented in Section~\ref{sec:results}. Finally, a summary and the
future prospects are given in Section~\ref{sec:summary}.

\section{Principle for vertex determination and prototype}
\label{sec:principle}
The present S$^4$AT prototype was designed so as to achieve depth resolution
below 0.29 mm (in sigma), or 1/$\sqrt{12}$ mm, for the reaction vertex. This
depth resolution is necessary to meet our requirement for excitation energy
resolutions of about 1 MeV in p($^6$He,d) and p($^6$Li,d) reaction measurements
in inverse kinematics using 400 -- 800 MeV/u secondary beams at GSI
Helmholtzzentrum für Schwerionenforschung GmbH, Darmstadt. The goal of the
experiment is to observe possible different effects of tensor interactions in
the ground states of $^6$He and $^6$Li~\cite{Horiuchi07}. In this section, we
describe the principle for vertex determination and the constructed prototype.

\subsection{Principle for vertex determination}
\begin{figure}[htbp] 
  \begin{centering}
    \includegraphics[scale=0.4]{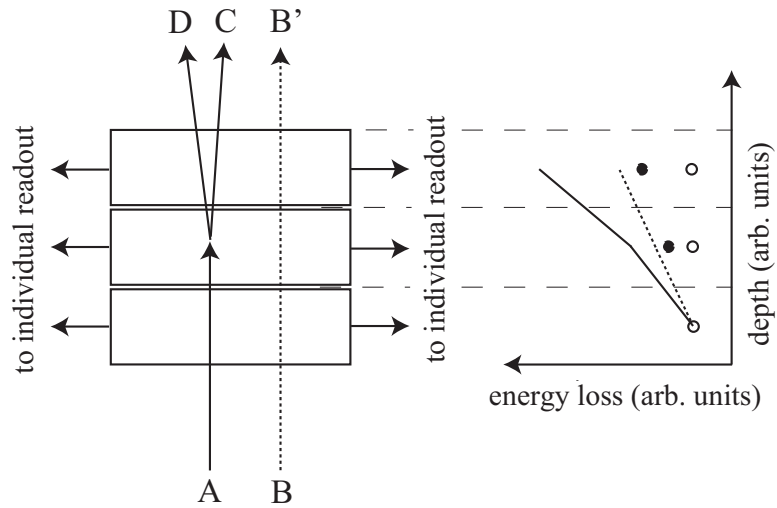}
    \caption{\label{fig:concept}Conceptual drawing of a three-layer
      plastic-scintillator active-target system. A and B represent two
      identical charged beam particles with the same kinetic energy. Whereas A
      reacts with the detector-target nucleus to produce charged particles C
      and D, B traverses the material medium without undergoing any nuclear
      reaction, losing its energy very gradually in the three plastic
      scintillators. For the A-induced event, however, the measured energy
      losses in the three layers are very different; the energy loss increases
      sharply in the second and third layers as compared to the case for the
      particle B. Right graph shows the energy loss (symbols) and cumulative
      energy loss (lines) of charged particles in an experiment with a $^6$He
      beam. The filled (open) symbol and solid (dotted) line correspond to the
      A (B)-induced event. See text for details.}
  \end{centering}
\end{figure}
%

The left diagram of Fig.~\ref{fig:concept} shows a conceptual drawing of a
three-layer plastic-scintillator active-target system. Here, we consider two
identical charged beam particles, denoted by A and B, with the same kinetic
energy. The particle A reacts with the detector-target nucleus to produce two
charged particles C and D. The particle B, on the other hand, traverses the
material medium without undergoing any nuclear reaction, losing its energy very
gradually in the three plastic scintillators. In the A-induced reactions,
however, the measured energy losses in the three layers are very different; the
energy loss increases sharply in the second and third layers as compared to the
case for the particle B. Hence, by measuring the energy loss in each of the
three layers, and considering the energy-loss distribution across the layers,
we can determine the reaction point in the thickness direction.

The present active target can be used in experiments with nuclear reactions in
both normal and inverse kinematics, although its usefulness in the former may be
limited. Some examples of the A-induced reactions in normal kinematics are
elastic proton-proton $^1$H(p,p) and proton-deuteron $^2$H(p,p) scatterings.
For reactions in inverse kinematics, a beam of the nucleus of interest, usually
with a kinetic energy of the order of several tens of MeV to several GeV,
bombards a target. The A-induced reaction then represents reactions in which a
beam-like and a target-like particles emerge from the target. Possible
reactions include elastic and inelastic scatterings off a proton or deuteron
target, transfer reactions such as (p,d), (d,p), (d,t) and (d,$^3$He),
charge-exchange reactions such as (p,n) and (d,2p), proton knockout reactions
and multi-nucleon transfer reactions.

To demonstrate the working principle of S$^4$AT, we consider the p($^6$He,d)
reaction induced by a 400-MeV/u $^6$He beam, and a layer thickness of 1 mm for
S$^4$AT. In this reaction, the incident $^6$He beam particle (particle A) is
stripped of a neutron by a target proton in S$^4$AT, leaving the residual ion
$^5$He (particle C) in its ground or excited state. The neutron and the proton
stick together to form a deuteron (particle D). Since $^5$He is particle
unbound, it decays almost immediately into a neutron and an $\alpha$ particle.
Compared to charged particles, neutrons interact very weakly with matter via
the electromagnetic interactions, and thus the measured energy loss in S$^4$AT
is mainly contributed by the deuteron and $\alpha$. For simplicity, we consider
reaction kinematics where both deuteron and $\alpha$ move in the beam direction.
Assuming that the reaction occurs at the centre of the second layer, the energy
of $^6$He right before the reaction, and those of deuteron and $\alpha$ right
after the reaction are 399.7 MeV/u, 31.7 MeV/u and 476.7 MeV/u, respectively.
The energy loss by the charged particles in each layer of the plastic
scintillators and the cumulative energy loss are shown by the filled symbols and
solid line on the right graph of Fig.\ref{fig:concept}, respectively. The open
symbols and dotted line represent the distributions for the non-reacted (B)
event. As will be shown in Section~\ref{sec:results}, for single-hit events,
the total energy loss (or even the energy loss in individual layers if the
energy resolution is sufficiently good) has a one-to-one correspondence with
the depth of reaction. The layer thickness of 1 mm was decided
based on a simple evaluation using a Monte Carlo simulation, and after taking
into account the energy losses and the energy resolution of the plastic
scintillators.

The one-to-one correspondence between energy loss and reaction depth may become
obscure in the presence of multi-hit events. The probability for multi-hit
events increases with increasing beam intensity. The admixture of the B-like
particles in the reaction channel of interest will alter the energy-loss
distributions, and may in some cases, degrade the energy resolution and
undermine its performance as an active target. Such an effect of non-reacted
particle(s) can be eliminated by considering the energy-loss difference between
adjacent layers. The depth information is reflected in the correlation
between energy-loss differences of two pairs of adjacent-layer combinations.
This method will be discussed in Section~\ref{sec:qrel}.

\subsection{S$^4$AT prototype}
\begin{figure}[htbp] 
  \begin{centering}
    \includegraphics[scale=0.4]{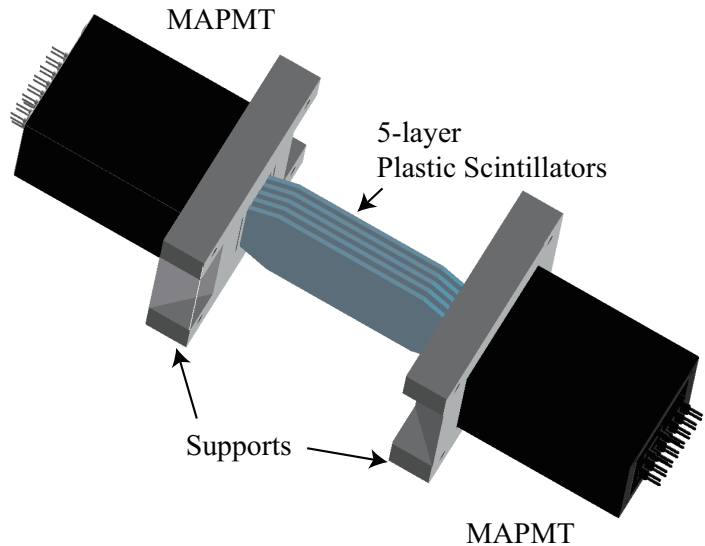}
    \caption{\label{fig:s4at}Schematic drawing of the S$^4$AT prototype.}
  \end{centering}
\end{figure}
We have constructed a prototype S$^4$AT system. Figure \ref{fig:s4at} shows a
schematic drawing of the S$^4$AT prototype. The prototype consists of five
1-mm-thick NE-102E-equivalent plastic scintillators, each with a sensitive area
of 50 (width) $\times$30 (height) mm$^2$. For better light collection and to
avoid light leakage, each plastic scintillator is wrapped with a 9-$\mu$m-thick
aluminized Mylar foil; the foil covers four surfaces which include the two
large-area surfaces of the scintillator. The unwrapped sides of the
scintillators are connected to two 16-channel Linear Array Multi-Anode
Photomultiplier Tube (MAPMT) assemblies H10515B-20 from Hamamatsu Photonics,
with which the light outputs are read out. Each anode has a 0.8$\times$16 mm$^2$
sensitive area, and the pitch size of the anodes is 1.0 mm. To prevent light
leakage from neighbouring scintillators, the plastic scintillators are arranged
in parallel at intervals of 1 mm. An opaque support made of Monomer-Cast (MC)
Nylon with rectangular holes, also at intervals of 1 mm and sufficient to allow
the plastic scintillators through, was mounted before the photocathode on each
side of the scintillators to fix their positions.

\section{Experiment}
\label{sec:exp}
\begin{figure}[htbp] 
  \begin{centering}
    \includegraphics[scale=0.4]{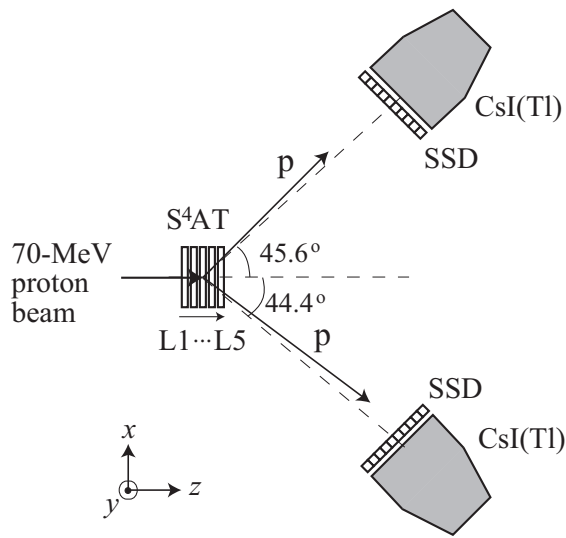}
    \caption{\label{fig:setup}Schematic layout of the experimental setup. The
      scale factor is arbitrary; the size of S$^4$AT and the thickness
      of silicon-strip detectors have been exaggeratedly enlarged for clarity.}
  \end{centering}
\end{figure}
The constructed S$^4$AT prototype was tested using elastic proton scattering
induced by a 70-MeV proton beam at the 41st beam line of CYRIC, Tohoku
University. This reaction was selected because of the similarity between the
energy losses of the proton beam and the scattered protons with those of the
charged particles in the p($^6$He,d) reaction with a 400-MeV/u $^6$He beam, as
well as the relatively large cross section and high luminosity.
Fig.~\ref{fig:setup} shows a schematic layout of the experimental setup. The
S$^4$AT system was mounted on a target ladder with the 50$\times$30-mm$^2$
surface perpendicular to the beam, and placed at the centre of a scattering
chamber. We define the axis along the beam direction as the $z$ axis; the $x$
axis is defined as shown in Fig.~\ref{fig:setup}, while the $y$ axis is the axis
emerging perpendicularly from the plane of the paper. The S$^4$AT layers from
upstream to downstream are denoted as L1 -- L5 according to their orders in the
beam direction. For convenience, we ignore the gap between layers, and define
the upstream surface of L1 and the downstream surface of L5 as $z$=0 and 5 mm,
respectively. In addition to S$^4$AT, we also mounted a 170-$\mu$m-thick
polyethylene (CH$_2$) target on the target ladder, and used it several times
during the experiment to monitor the stability of the beam-energy spread and
detector resolution.

To evaluate the depth resolution of S$^4$AT and to enable selection of the
elastic scattering channel, we placed two sets of Si-CsI(Tl) telescopes at
around 45$^\circ$ with respect to the incident proton beam. The Si-CsI(Tl)
telescope with an active area of 50$\times$50 mm$^2$ consists of a
300-$\mu$m-thick single-sided silicon-strip detector (SSD) with 10 strips from
Hamamatsu Photonics and a 55-mm-thick thallium-doped caesium iodide
scintillation crystal from Institute of Modern Physics, Chinese Academy of
Sciences. The distance from the target position to the SSDs was about 30 cm.
The proton scattering angles were determined from their hit positions on the
SSDs, while the kinetic energies were determined from the energy losses in SSDs
and CsI(Tl) detectors. The angle subtended by one strip of the SSD, with a
5-mm width, is about 1$^\circ$. The rear part of the CsI(Tl) crystal (with a
thickness of 30 mm) is cut into a trapezoid shape with a reduced area of 30 mm
$\times$ 30 mm, and is attached to a Si-PIN photodiode S3584-08 from Hamamatsu
Photonics. The charge signals from the SSDs and Si-PIN photodiodes were
collected and processed by 16-channel charge-sensitive preamplifiers MPR-16
(from Mesytec GmbH \& Co) and KPA-16 (from Kaizuworks), respectively. The
output signals were later processed by Mesytec's MSCF-16 shaping amplifiers
before being fed to 32-channel peak-sensing analog-to-digital converters
(MADC-32). 

For S$^4$AT, the output signals from MAPMT were sent to a photomultiplier
amplifier. One of the two outputs from each channel was sent through a cable
delay before being fed into a 32-channel charge-to-digital converter (CAEN V792)
module; the other output was fed into a leading-edge discriminator to generate
a timing signal, which was mainly used to determine the beam intensity.

Measurements were performed using proton beams with intensities ranging from 
about 75 kcps to 3.6 Mcps. The beam intensities were determined combining the
scaler information for S$^4$AT and the telescopes, as well as the relative
event numbers from measurements with different beam intensities. For the
measurements with the CH$_2$ target, a beam intensity of about 0.8 nA was used.
Data acquisition was performed with the logic OR signal of the SSD signals as
trigger using the software package babirlDAQ~\cite{Baba10}.

\section{Data analysis and results}
\label{sec:results}
\begin{figure}[htbp] 
  \begin{centering}
    \includegraphics[scale=0.4]{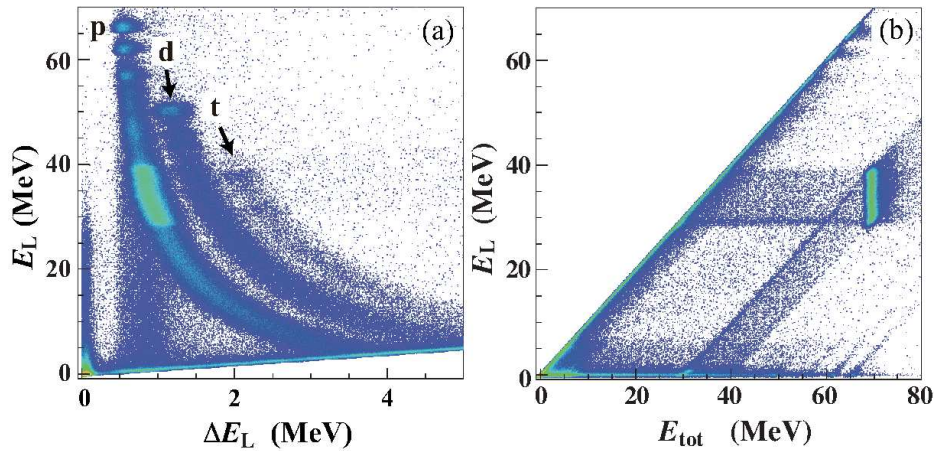}
    \caption{\label{fig:deech2} Energy loss ($\Delta E$) and total
      energy loss ($E$) of charged particles in the SSD and SSD-CsI(Tl)
      detectors during the measurement with the CH$_2$ target. (a) A scatter
      plot showing the correlation between the measured $E_{\rm L}$ and
      $\Delta E_{\rm L}$ in the left Si-CsI(Tl) telescope. (b) Correlation
      between the total energy loss measured by the left telescope ($E_{\rm L}$)
      and the summed energy loss in the left and right telescopes
      ($E_{\rm tot} \equiv E_{\rm L}+E_{\rm R}$). The dense locus corresponds to
      the proton-proton elastic scattering events.}
  \end{centering}
\end{figure}
\begin{figure}[htbp] 
  \begin{centering}
    \includegraphics[scale=0.4]{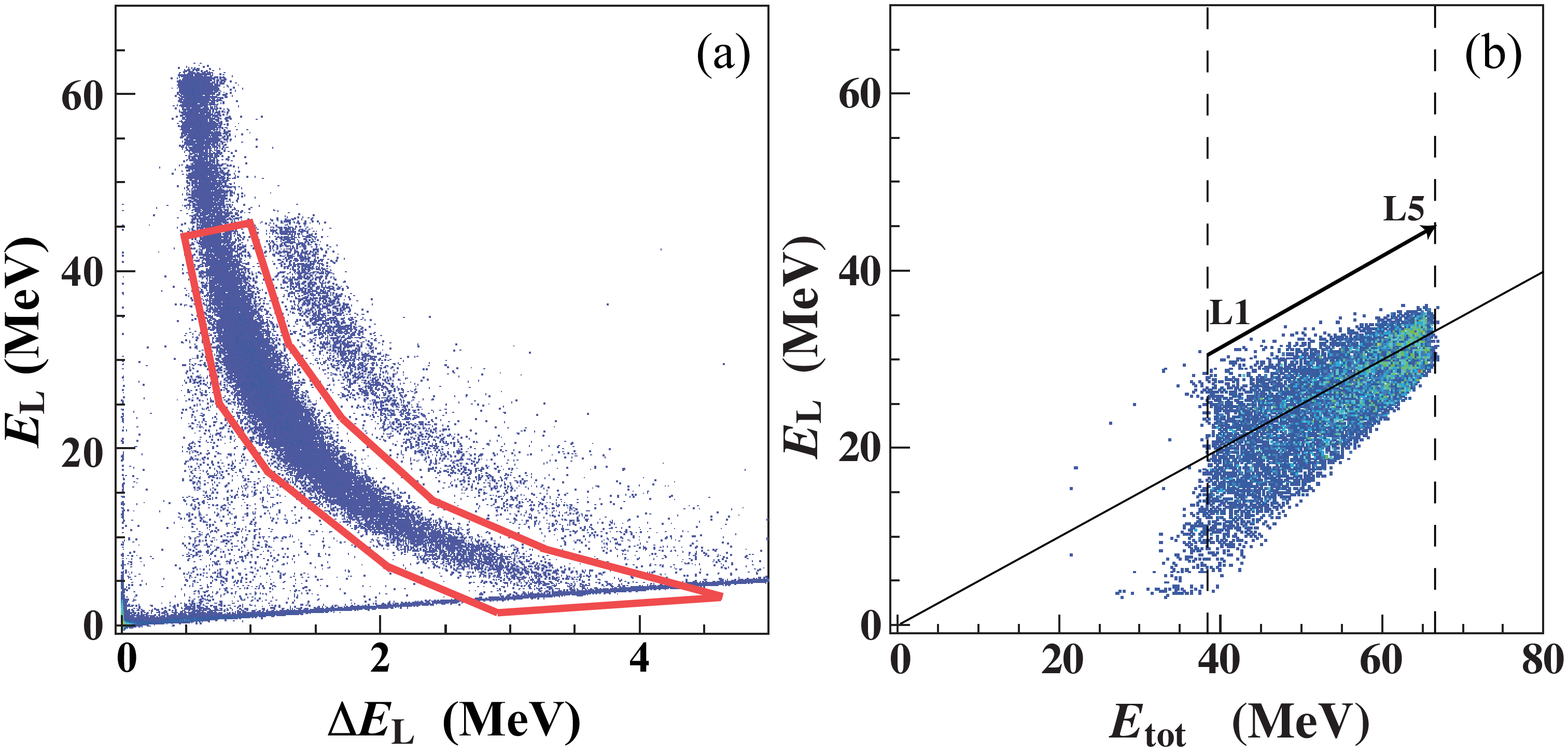}
    \caption{\label{fig:deeat} (a) Scatter plot from the measurement with
      S$^4$AT at 75-kcps beam intensity showing the correlations between the
      measured $E_{\rm L}$ and $\Delta E_{\rm L}$ with the left Si-CsI(Tl)
      telescope. The red polygon indicates the typical gate used to select
      elastically-scattered protons. (b) The total energy loss measured by the
      left ($E_{\rm L}$) and the summed energy loss in the left and right
      telescopes ($E_{\rm tot}$) after applying the proton gates from the left
      and right telescopes. The black arrow in (b) indicates roughly the depth
      position of the reaction vertex in the beam direction (L1 -- L5). The
      vertical dashed lines in the figure indicate the positions
      of $z$=0 and 5 mm, and are only a guide for the eye. The inclined solid
      line is the equal-energy line which represents the elastically-scattered
      protons with the same scattering angle. 
    }
  \end{centering}
\end{figure}
The data analyses were performed using the object-oriented data analysis
framework, ROOT~\cite{ROOT}. Fig.~\ref{fig:deech2}(a) shows the correlation
between the energy loss ($\Delta E_{\rm L}$) and total energy loss ($E_{\rm L}$)
of charged particles in the left SSD and the left SSD-CsI(Tl) telescope for the
measurement with the CH$_2$ target. Three loci corresponding to protons,
deuterons and tritons are observed. These particles originated from elastic and
inelastic proton scatterings on proton (the dense elongated locus) and $^{12}$C,
and (p,d) and (p,t) reactions on $^{12}$C, respectively. The small clusters
distributed along the proton and deuteron loci correspond to the ground and
excited states in $^{12}$C and $^{11}$C. These states, together with the
proton-proton elastic scattering events measured at different angles, were used
for the energy calibration of the SSD and CsI(Tl) detectors.
Fig.~\ref{fig:deech2}(b) shows the correlation between the total energy losses
measured by the left telescope ($E_{\rm L}$) and the summed energy loss in the
left and right telescopes ($E_{\rm tot} \equiv E_{\rm L}+E_{\rm R}$). The dense
locus corresponds to proton-proton elastic scattering events.

The energy resolution of the telescopes for proton scattering was evaluated
using the observed discrete states in $^{12}$C as well as the proton elastic
scattering events during the measurement with the CH$_2$ target. Since the
energy resolutions of the SSDs were below 0.2\% for 1-MeV energy loss of
protons, and the elastically-scattered protons only deposited less than 5\% of
their energies in the SSDs compared to the CsI(Tl), we considered only the
resolutions of the CsI(Tl) detectors. The widths of the observed discrete
states can be attributed to the beam-energy spread, the effect of the target
thickness, which includes reaction depth dependence and energy straggling,
energy spread due to the reaction kinematics, and the energy resolutions of the
CsI(Tl) detectors. To understand the detector response, we performed Monte Carlo
simulations with the GEANT4 toolkit~\cite{Geant4}, taking into account the
experimental setup, reaction kinematics, beam-energy spread and intrinsic
energy resolutions of the detectors. The energy spread due to the target
thickness contributes only about 7\% (in terms of squared relative uncertainty)
to the observed width, and is thus negligible. From comparison between the
simulations and the experimental data, the beam-energy spread was found to be
less than 200 keV (in sigma). The relative energy resolutions of the CsI(Tl)
detectors thus determined consist of three terms~\cite{Brown12}, namely the
stochastic, electronic-noise and constant terms as shown below:
\begin{equation}
  \label{eq_res}
  \frac{\sigma(E)}{E} = \frac{k_{\rm stochastic}}{\sqrt{E}} + \frac{k_{\rm electronic}}{E} + k_{\rm constant}.
\end{equation}
$k_{\rm stochastic}$, $k_{\rm electronic}$ and $k_{\rm constant}$ are the coefficients
with typical values of -0.61 MeV$^{1/2}$, +2.55 MeV and +0.04, respectively,
assuming a 200-keV beam-energy spread. This relation was later used in the
simulations for the measurements with S$^4$AT.

Figure~\ref{fig:deeat}(a) shows similar $E_{\rm L}$-$\Delta E_{\rm L}$ plot but
for the measurement with S$^4$AT using a 75-kcps proton beam. The red polygon
indicates the typical gate used to select elastically-scattered protons. A
similar gate was also defined for the right SSD-CsI(Tl) telescope, and the
combined gates were taken as the `pp-elastic' proton gates.
Figure~\ref{fig:deeat}(b) shows the $E_{\rm L}$ versus $E_{\rm tot}$ plot for the
measurement with S$^4$AT after applying the `pp-elastic' proton gates. The
$E_{\rm L}$-$E_{\rm tot}$ plot is more widely distributed compared to the case
for the CH$_2$ target. The black arrow indicates roughly the depth position of
the reaction vertex in the beam direction (L1 -- L5). The vertical dashed lines
in the figure indicate the positions of $z$=0 and 5 mm, and are only a guide
for the eye. The inclined solid line is the equal-energy line which represents
the elastically-scattered protons with the same scattering angle. Protons with
larger (smaller) scattering angles are distributed below (above) the
equal-energy line. The energy spread is larger for reactions in L1 than in L5,
and for larger scattering angles than for smaller scattering angles, because
the scattered protons had to penetrate more materials before reaching the
telescopes, and therefore are subjected to more energy-loss straggling and
multiple Coulomb scattering.


To determine the depth of the reaction vertex, we apply three methods using the
following information: (i) the total energy loss of the scattered protons in
the telescopes, (ii) the total energy loss of charged particles in S$^4$AT, and
(iii) energy-loss differences between different layers of S$^4$AT.

\subsection{Vertex determination with SSD-CsI(Tl) telescope}
\label{sec:res1}
\begin{figure}[htbp] 
  \begin{centering}
    \includegraphics[scale=0.4]{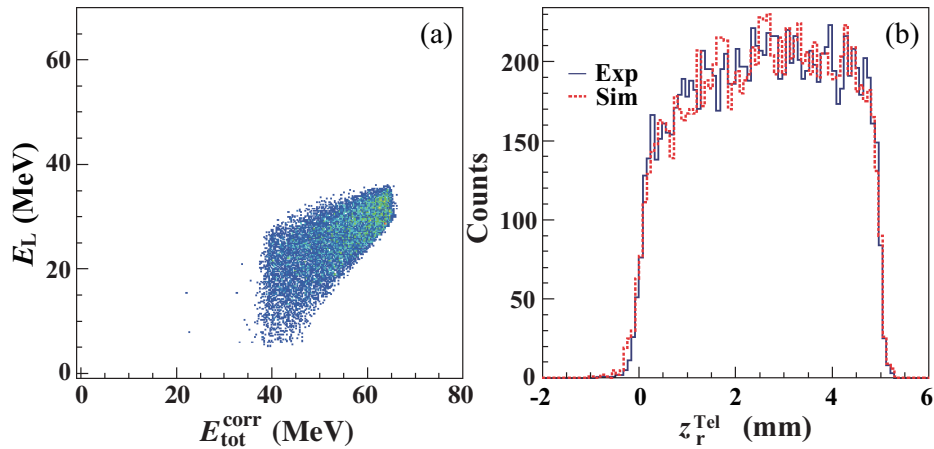}
    \caption{\label{fig:zdee} (a) Total energy loss in the left telescope
      ($E_{\rm L}$) against the corrected summed energy loss
      ($E^{\rm corr}_{\rm tot}$) for the measurement with
      S$^4$AT at 75-kcps beam intensity. (b) Reaction vertex distribution
      reconstructed from the total proton energies measured with the left and
      right SSD-CsI(Tl) telescopes. Only pp-elastic protons
      have been selected.
      The red-dotted histogram shows the results of the Monte Carlo simulation.}
  \end{centering}
\end{figure}
\begin{figure}[htbp] 
  \begin{centering}
    \includegraphics[scale=0.4]{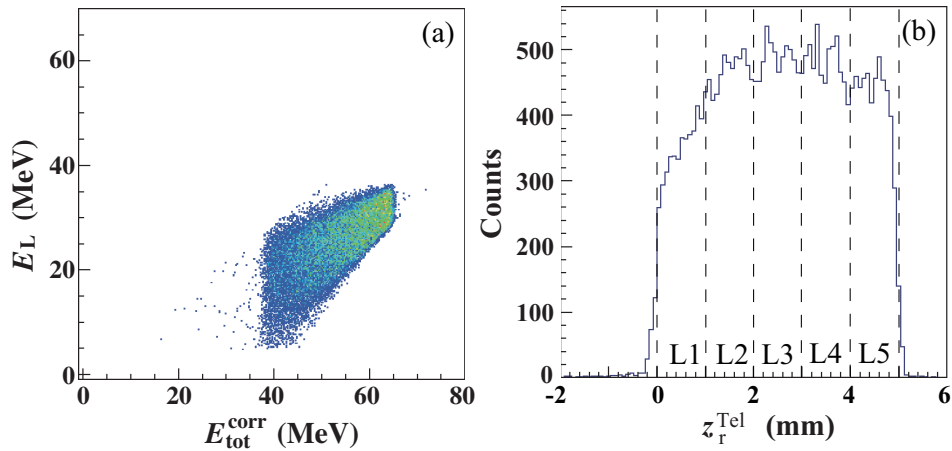}
    \caption{\label{fig:zdee2} (a) Total energy loss in the left telescope
      ($E_{\rm L}$) against the corrected summed energy loss
      ($E^{\rm corr}_{\rm tot}$), and (b) reconstructed depth distribution for the
      measurement with 3.6-Mcps proton beam.}
  \end{centering}
\end{figure}
The depth of the reaction vertex was first determined using the summed kinetic
energy of the pp-elastic protons measured by the two SSD-CsI(Tl) telescopes.
For a reaction at a depth position $z_{\rm r}$, the total kinetic energy of the
scattered protons (denoted by p$_{\rm L}$ and p$_{\rm R}$) right after the
reaction, denoted by $E_{\rm tot}(z=z_{\rm r})$, is equal to the kinetic energy of
the proton beam right before the reaction $E_{\rm B}(z_{\rm r})$.
$E_{\rm B}(z_{\rm r})$ is unique if we assume a parallel proton beam. The kinetic
energies of p$_{\rm L}$ [$E_{\rm L}$($z_{\rm r};\theta_{\rm L},z=z_{\rm r}$)] and
p$_{\rm R}$ [$E_{\rm R}$($z_{\rm r};\theta_{\rm R},z=z_{\rm r}$)] depend on
their scattering angles ($\theta_{\rm L}$ and $\theta_{\rm R}$). p$_{\rm L}$ and
p$_{\rm R}$ penetrated the S$^4$AT detector, losing part of their energies before
being stopped and detected by the SSD-CsI(Tl) telescopes. For simplicity, the
kinetic energies of the protons measured by the left and right telescopes are
denoted by $E_{\rm L}$ and $E_{\rm R}$, respectively. In principle,
$E_{\rm L}+E_{\rm R}$ ($\equiv E_{\rm tot}$) should have a one-to-one
correspondence with the reaction depth. However, as can be seen in
Fig.~\ref{fig:deeat}(b), the $E_{\rm tot}$ distribution is skewed especially for
protons with larger scattering angles (smaller $E_{\rm L}$) due to their larger
energy spread. This skewing effect was estimated using Monte Carlo simulations,
and $E_{\rm tot}$ was corrected for each $E_{\rm L}$ value. The corrected
distribution ($E^{\rm corr}_{\rm tot}$) is shown in Fig.~\ref{fig:zdee}(a). In
this way, we obtained a linear conversion function between the total energy of
the scattered protons and the reaction depth.  The reconstructed depth
distribution is shown in Fig.~\ref{fig:zdee}(b). For comparison, the depth
distribution reconstructed from Monte Carlo simulation data is also shown (the
red-dotted histogram). Overall, the simulation reproduced the experimental data
very well.

%
%
It is important to note that the resolution of the reaction depth determined
with the telescopes remained almost the same up to 3.6-Mcps proton beam, as
shown in Fig.~\ref{fig:zdee2}. The detected events reacted in L1, which include
the events at $z < 0$ mm, are about
16\% less compared to other layers, due likely to reduced acceptance of the
telescopes for protons with large scattering angles.
These missing events correspond to about 2\% of the expected total pp-elastic
events assuming that the elastic scattering cross section remains constant in
all layers. 

\subsection{Vertex determination with S$^4$AT: Total energy-loss method}
\label{sec:qsum}
\begin{figure}[htbp] 
  \begin{centering}
    \includegraphics[scale=0.35]{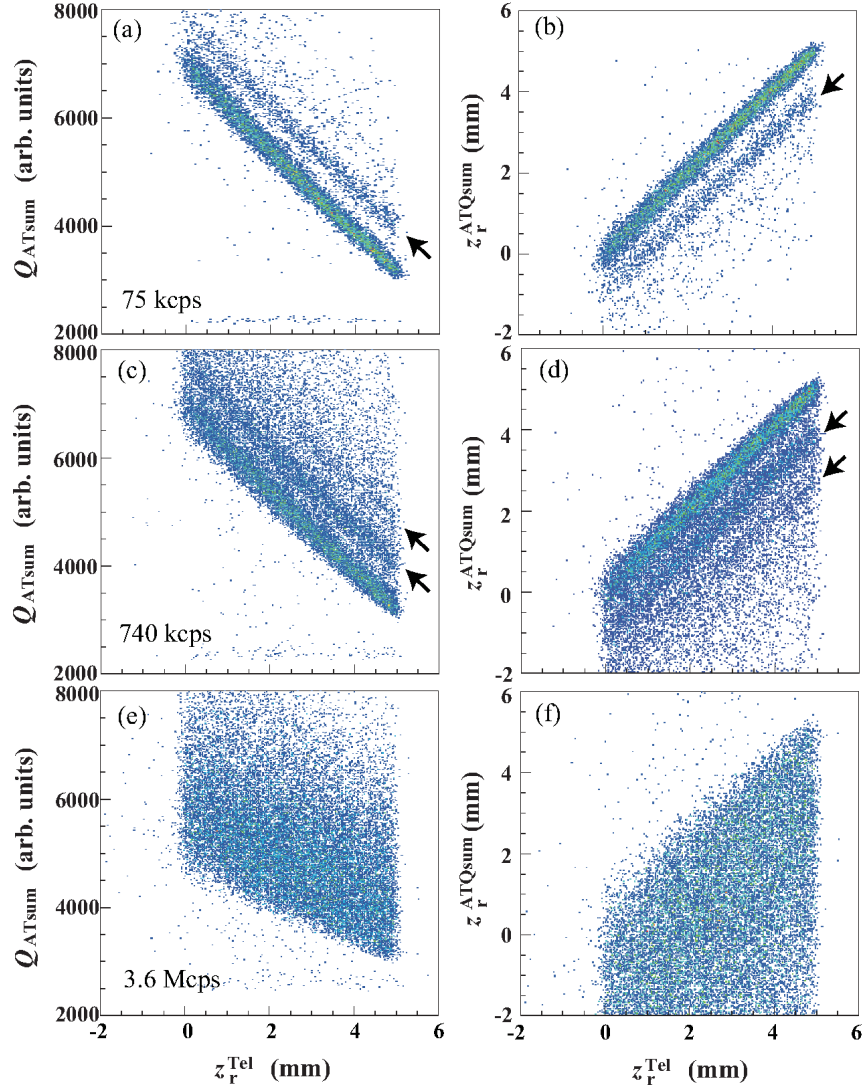}
    \caption{\label{fig:atsum} (a)(c)(e) Total energy loss in S$^4$AT versus
      the reaction depth determined using the SSD-CsI(Tl) telescopes
      $z^{\rm Tel}_{\rm r}$, and (b)(d)(f) reaction depth reconstructed
      from the total energy loss in S$^4$AT, $z^{\rm ATQsum}_{\rm r}$ versus
      $z^{\rm Tel}_{\rm r}$. (a) and (b), (c) and (d), (e) and (f) are for
      measurements with 75-kcps, 740-kcps and 3.6-Mcps beams, respectively.
      The arrows in (a) -- (d) indicate shifts due to pile-ups.}
  \end{centering}
\end{figure}
The depth of the reaction vertex can be determined by simply considering
cumulative (total) energy loss of charged particles in S$^4$AT. This is
equivalent to replacing the five layers by a single 5-mm-thick plastic
scintillator. In this method, the total energy loss of the proton beam before
reaction and the two protons (p$_{\rm L}$ and p$_{\rm R}$) after reaction in
S$^4$AT is measured. The total energy loss in S$^4$AT decreases with
increasing $z_{\rm r}$ because p$_{\rm L}$ and p$_{\rm R}$, which have lower
energies, penetrate through thinner plastic scintillator material.
Fig.~\ref{fig:atsum}(a) shows the correlation between total energy loss
(denoted by $Q_{\rm ATsum}$) and the reaction depth determined using the
SSD-CsI(Tl) telescopes as described in Section~\ref{sec:res1}. Here, only events
within the `pp-elastic gates' have been selected. Events due to pile-up, which
manifest in the form of energy-loss shifts as shown by the arrows, in S$^4$AT
are clearly observed. The energy resolution for 1-MeV energy loss of a proton
beam in a layer of S$^4$AT was about 18\% at 75-kcps beam intensity.
As will be discussed in Section~\ref{sec:resoeff}, the reaction depth
resolution depends on the energy resolution of S$^4$AT.

To reconstruct the reaction depth from $Q_{\rm ATsum}$, we derived a conversion
formula using the single-hit events in Fig.~\ref{fig:atsum}(a). The resulting
reaction depth (denoted by $z^{\rm ATQsum}_{\rm r}$) is plotted against $z_{\rm r}$
determined by the telescopes (denoted by $z^{\rm Tel}_{\rm r}$) in
Fig.~\ref{fig:atsum}(b). For comparison, similar plots for the measurement with
740-kcps proton beam are shown in Figs.~\ref{fig:atsum}(c) and (d). The energy
resolution deteriorates due to the increased pile-up events. It is important to
note that although the single-hit, double-hit and even the triple-hit loci are
still distinguishable, without the depth information from the telescope, it
will be difficult to efficiently determine the reaction depth. The condition
worsened at 3.6-Mcps beam intensity as shown in Figs.~\ref{fig:atsum}(e) and
(f). Besides the deterioration in the energy and reaction depth resolutions, one
observes, by comparing Fig.~\ref{fig:atsum}(e) with those of (c) and (a), clear
reduction in the pulse heights of all anode signals. Such reduction is due to
possible increased electron current which results in a drop in the interstage
voltages for the dynodes in the MAPMTs, and consequently reduced collection
efficiency and secondary electron emission. Although the gain-quenching effect
can be addressed by modifying the last few stages of the
MAPMTs to allow additional power supply boosters, the pile-up events will
represent the major challenge for this method at a high-counting rate. In the
next section, we demonstrate a novel method to address this issue.

\subsection{Vertex determination with S$^4$AT: Energy-loss difference method}
\label{sec:qrel}
\begin{figure}[htbp] 
  \begin{centering}
    \includegraphics[scale=0.4]{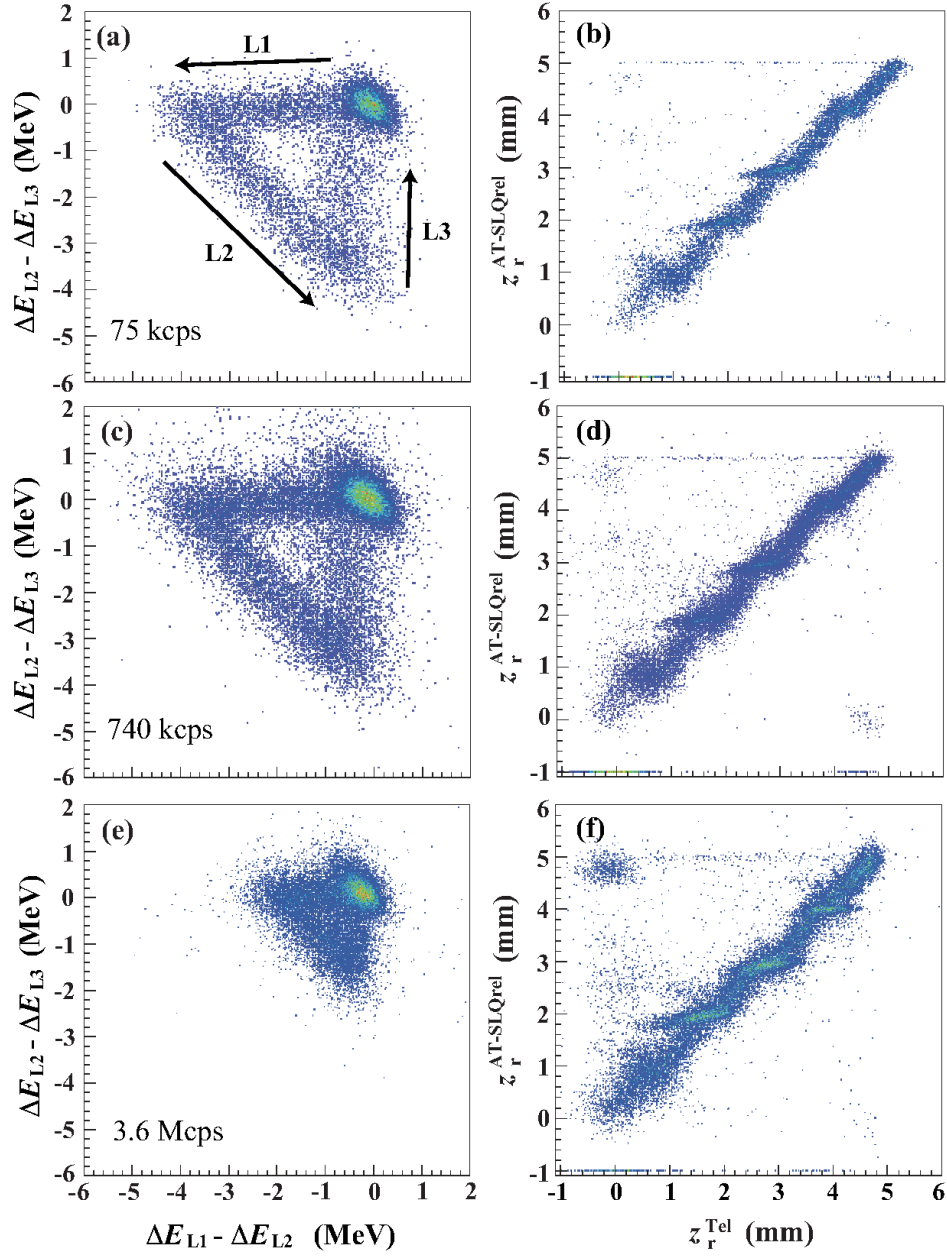}
    \caption{\label{fig:atdiff} (a) Energy-loss difference between L2 and L3
      ($\Delta E_{\rm L3}-\Delta E_{\rm L2}$) versus energy-loss difference
      between L1 and L2 ($\Delta E_{\rm L2}-\Delta E_{\rm L1}$). The arrows
      indicate the direction of the depth position from L1 to L3. (b) Reaction
      depth, $z^{\rm AT-SLQrel}_{\rm r}$, reconstructed from single-layer
      energy-loss-difference method versus $z^{\rm Tel}_{\rm r}$. (c) -- (d)
      and (e) -- (f) are similar plots for the measurements with 740-kcps and
      3.6-Mcps beams, respectively.}
  \end{centering}
\end{figure}
\begin{figure}[htbp] 
  \begin{centering}
    \includegraphics[scale=0.4]{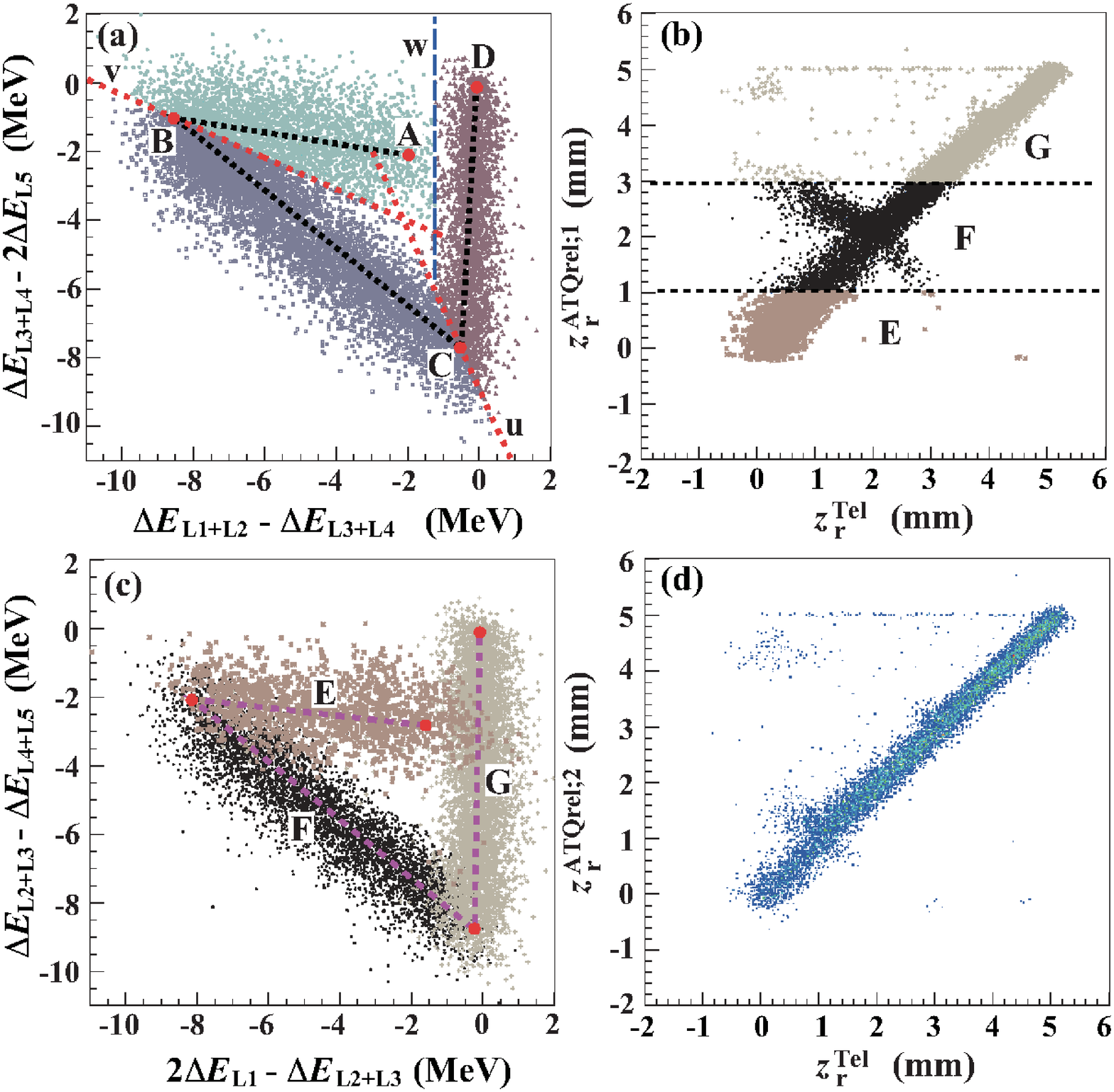}
    \caption{\label{fig:atdiff2} (a) Correlation between $\Delta E_{\rm L1+L2} -
      \Delta E_{\rm L3+L4}$ and $\Delta E_{\rm L3+L4} - 2\Delta E_{\rm L5}$.
      (b) Reaction depth, denoted as $z^{{\rm ATQrel};1}_{\rm r}$, reconstructed
      from the multi-layer energy-loss-difference correlation shown in (a)
      versus $z^{\rm Tel}_{\rm r}$. (c) Correlation between $\Delta E_{\rm L2+L3}
      - \Delta E_{\rm L4+L5}$ and $\Delta E_{\rm L4+L5} - 2\Delta E_{\rm L1}$. The
      coloured events correspond to the events in the three regions separated
      by the two lines shown in (b). (d) Reaction depth, denoted as
      $z^{{\rm ATQrel};2}_{\rm r}$, reconstructed from the multi-layer
      energy-loss-difference correlation shown in (c) versus
      $z^{\rm Tel}_{\rm r}$. The data are from measurements with 75-kcps proton
      beam.}
  \end{centering}
\end{figure}
As mentioned earlier, the presence of nonreacted beam particle(s) in a reaction
event results in a shift in the measured energy loss. For intermediate-energy
beam passing through thin layers of plastic scintillators, the energy loss of a
beam particle in each layer is almost constant. Hence, this energy-loss shift,
i.e. the effect of pile-up, can be eliminated by taking the energy-loss
difference between adjacent layers. The simplest method is to consider energy
loss in every single layer. Fig. \ref{fig:atdiff}(a) shows the scatter plot for
the energy-loss difference between L2 and L3
($\Delta E_{\rm L3}-\Delta E_{\rm L2}$) against the energy-loss difference
between L1 and L2 ($\Delta E_{\rm L2}-\Delta E_{\rm L1}$) for the measurement with
75-kcps proton beam. Here, we have selected only coincident protons using
the telescopes. Different points on the sides of the triangular locus
correspond to different reaction depth in layers L1 -- L3. The arrows indicate
the direction of the depth position. The dense elliptical events correspond to
events reacted in L4 and L5. Similar correlations can be obtained by
considering other combinations of energy-loss difference, and used to
determine the reaction depth. This single-layer energy-loss-difference method,
however, is subject to inherent uncertainties around the borders of adjacent
layers, which correspond to the corners of the triangular locus in
Fig. \ref{fig:atdiff}(a). For comparison, we plot the reaction depth
reconstructed from this method, denoted as $z^{\rm AT-SLQrel}_{\rm r}$ against
$z^{\rm Tel}_{\rm r}$ in Fig.~\ref{fig:atdiff}(b). Figs.~\ref{fig:atdiff}(c) and
(d) show similar plots as (a) and (b) but for the measurement with the 740-kcps
beam. The loci in Fig.~\ref{fig:atdiff}(c) are almost identical to those in
Fig.~\ref{fig:atdiff}(a), although the resolution looks slightly worse. Similar
plots for the measurement with the 3.6-Mcps beam are shown in
Figs. \ref{fig:atdiff}(e) and (f). Although the energy-loss differences shrink
due to gain shift of the MAPMTs, it is still possible to determine the reaction
depth by applying a universal scaling since the gain shift is common for all
anodes. However, the uncertainties around the borders of adjacent layers and
consequently, the depth resolutions deteriorate due to the worsening energy
resolutions of the scintillators.

As a counter measure, we consider the correlation between
$(\Delta E_{\rm L1} + \Delta E_{\rm L2}) - (\Delta E_{\rm L3}+\Delta E_{\rm L4})$
and $(\Delta E_{\rm L3}+\Delta E_{\rm L4}) - (2\Delta E_{\rm L5})$. For simplicity,
we rewrite $\Delta E_{\rm L1} + \Delta E_{\rm L2} = \Delta E_{\rm L1+L2}$ and
$\Delta E_{\rm L3} + \Delta E_{\rm L4} = \Delta E_{\rm L3+L4}$. As an example,
we show the correlation for the measurement with 75-kcps proton beam in
Fig.~\ref{fig:atdiff2}(a). Notice that the L1-L2 (or L3-L4) border now lies on
the straight locus AB (BC), which forms one of the sides of the open triangle.
Besides addressing the border issue, summing the energy losses in two layers
also helps to improve the energy resolution. This method, called the multi-layer
energy-loss-difference method, consists of two stages of procedures. In the
first stage, we used the correlation in Fig. \ref{fig:atdiff2}(a) to determine
the reaction depth. The procedures in the first stage are described as follows:
\begin{enumerate}
\item Determine central lines of the loci along L1 -- L2, L3 -- L4 and L5, as
  shown by the dashed lines in Fig. \ref{fig:atdiff2}(a).
\item Determine the interior bisectors, lines denoted by ``u'' and ``v'', of
  the angles formed by ABC and BCD, as well as line ``w''; the line ``w'' is
  parallel to the vertical axis, and corresponds to the minimum position of the
  projected $\Delta E_{\rm L1+L2} - \Delta E_{\rm L3+L4}$ distribution.
\item Divide the scatter plot into three straight loci using combinations of
  lines.
\item For each straight locus, determine a linear conversion formula for the
  reaction depth.
\end{enumerate}
The reconstructed reaction depth, denoted as $z^{{\rm ATQrel};1}_{\rm r}$, is
plotted against $z^{\rm Tel}_{\rm r}$ in Fig. \ref{fig:atdiff2}(b). As shown in
the figure, the uncertainties in the reconstructed reaction depth are relatively
large near the borders of L2 -- L3 and L4 -- L5.

\begin{figure}[htbp] 
  \begin{centering} 
    \includegraphics[scale=0.4]{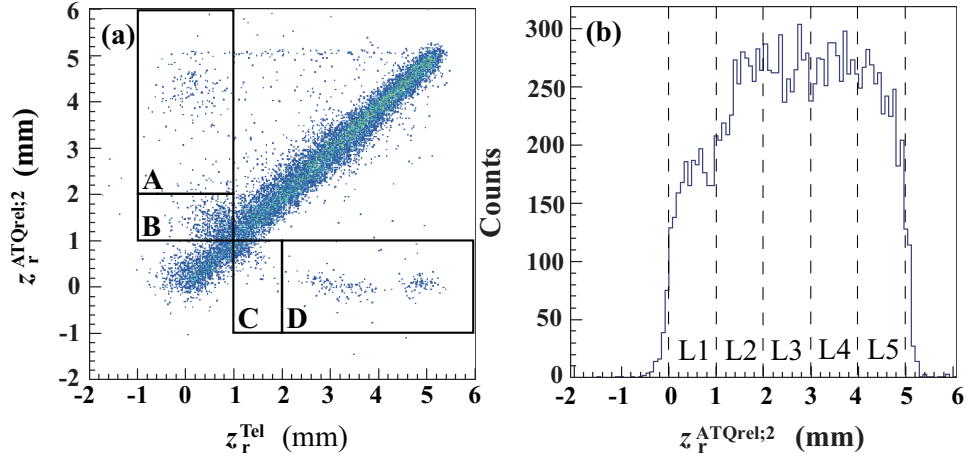}
    \caption{\label{fig:zat} (a) Reaction depth $z^{{\rm ATQrel};2}_{\rm r}$ versus
      $z^{\rm Tel}_{\rm r}$, and (b) $z^{{\rm ATQrel};2}_{\rm r}$ distribution for
      the measurement with 1.6-Mcps proton beam.}
  \end{centering}
\end{figure}
%
%
%
To reduce the uncertainties around the borders and improve the reaction depth
determination, we consider, in the second stage, the correlation between
$(\Delta E_{\rm L3} + \Delta E_{\rm L2}) - (\Delta E_{\rm L5}+\Delta E_{\rm L4})$
and $(2\Delta E_{\rm L1}) - (\Delta E_{\rm L5} + \Delta E_{\rm L4})$, as shown in
Fig.~\ref{fig:atdiff2}(c). By dividing $z^{{\rm ATQrel};1}_{\rm r}$ into three
regions, as indicated by E, F and G in Fig.~\ref{fig:atdiff2}(b), and selecting
each region, three straight loci corresponding to the reaction depth in
L1, L2 -- L3 and L4 -- L5 can be obtained. These straight loci were
then used to determine the reaction depth, $z^{{\rm ATQrel};2}_{\rm r}$. The
reaction depth thus obtained is plotted against $z^{\rm Tel}_{\rm r}$ in
Fig.~\ref{fig:atdiff2}(d).

We applied this method to measurements with various beam intensities.
Fig.~\ref{fig:zat} shows (a) the correlation between the reconstructed reaction
depths using S$^4$AT ($z^{{\rm ATQrel};2}_{\rm r}$) and the telescopes
($z^{\rm Tel}_{\rm r}$), and (b) the projected $z^{{\rm ATQrel};2}_{\rm r}$
distribution for the measurement with 1.6-Mcps proton beam. The dashed lines in
Fig.~\ref{fig:zat}(b) have been added for reference. The reacted events in L1,
which include also the events below 0 mm in Fig.~\ref{fig:zat}(b), are about
28\% less than in other layers.
As mentioned in Section \ref{sec:res1}, 16\% of the pp-elastic protons from L1
failed to reach the telescopes. The other 12\% missing events are fairly
accounted for by mis-reconstructions of reaction depth as shown by the four
components in Fig.\ref{fig:zat}(a): missing A (-6\%) and B (-20\%)
components, and additional C (+9\%) and D (+6\%) components. These
mis-reconstructions are due to the unresolved events in L1 and L5 around the
vertical ``w'' line in Fig.~\ref{fig:atdiff2}(a) (components A and D), and in
L1 and L2 around the intersection of the lines E and F in
Fig.~\ref{fig:atdiff2}(c) (components B and C). Such unresolved events can be
reduced by improving the energy resolution of the S$^4$AT system.

\subsection{Resolution and tracking efficiency}
\label{sec:resoeff}

\begin{figure}[h] 
    \includegraphics[scale=0.4]{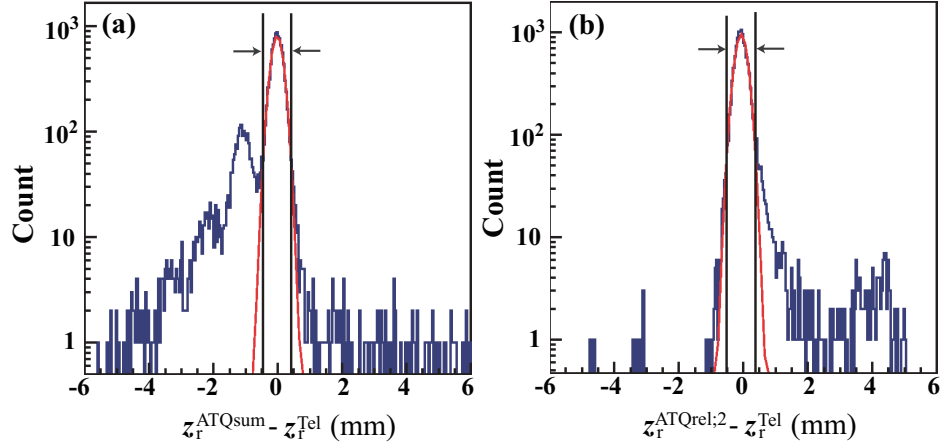}
    \caption{\label{fig:dz} $\Delta z^{\rm AT} \equiv z^{\rm AT}_{\rm r} - z^{\rm Tel}_{\rm r}$
      distributions for (a) $z^{\rm AT}_{\rm r} = z^{\rm ATQsum}_{\rm r}$ and
      (b) $z^{\rm AT}_{\rm r} = z^{{\rm ATQrel};2}_{\rm r}$ for the measurements with
      1.6-Mcps proton beam. The solid-vertical lines with arrows indicate the
      ranges for $|\Delta z^{\rm AT}| \leq 2.33\sigma_z^{\rm AT}$, where
      $\sigma_z^{\rm AT}$ is the root-mean-square width of the $\Delta z^{\rm AT}$
      distribution around 0.}
\end{figure}

\begin{figure}[htbp] 
  \begin{centering}
    \includegraphics[scale=0.4]{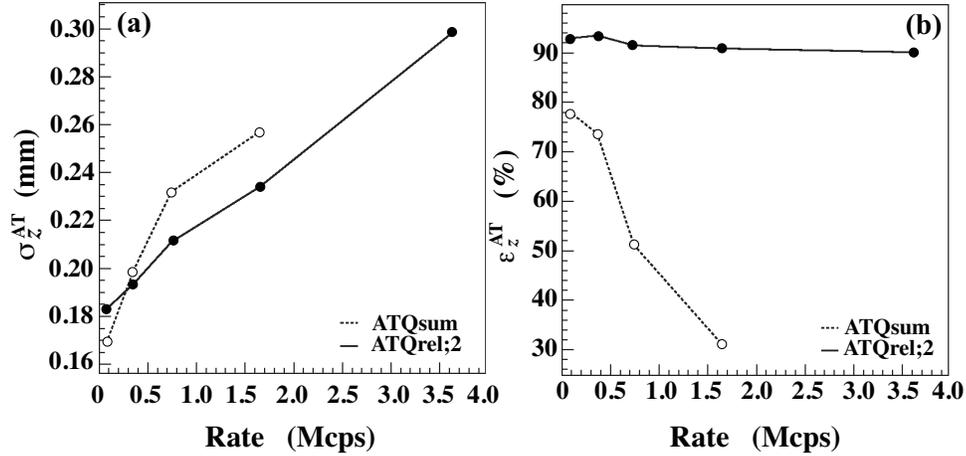}
    \caption{\label{fig:resoeffi} Beam-intensity dependence of (a) reaction
      depth resolution, and (b) tracking efficiency.}
  \end{centering}
\end{figure}
The performance of S$^4$AT is evaluated in terms of the depth position
resolution and tracking efficiency. To evaluate the depth position resolution,
we consider the difference between the reaction depth determined using the
telescopes and the one with S$^4$AT:
\begin{equation}
  \Delta z^{\rm AT} \equiv z^{\rm AT}_{\rm r} - z^{\rm Tel}_{\rm r}, 
\end{equation}
where AT is ``ATQsum'' or ``ATQrel;2''. The $\Delta z^{\rm AT}$ distributions are
shown in Figs.~\ref{fig:dz}(a) and (b). The root-mean-square width of the
$\Delta z^{\rm AT}$ distribution around 0, denoted as $\sigma_z^{\rm AT}$, is
taken as the effective depth position resolution, which is attributed to the
intrinsic energy resolutions of the plastic scintillators and the telescopes.
The tracking efficiency is defined as
\begin{equation}
  \epsilon_z^{\rm AT} \equiv N(|\Delta z^{\rm AT}| \leq 2.33\sigma_z^{\rm AT})
  /N(-1<z^{\rm Tel}_{\rm r}<6), 
\end{equation}
where $N(|\Delta z^{\rm AT}| \leq 2.33\sigma_z^{\rm AT})$ represents the number of
events with $\Delta z^{\rm AT}$ within $2.33\sigma_z^{\rm AT}$ or 98\%
coincidence level, and $N(-1<z^{\rm Tel}_{\rm r}<6)$ the number of events
successfully tracked by the telescopes. The dependences of $\sigma_z^{\rm AT}$
and $\epsilon_z^{\rm AT}$ on the beam intensity are plotted in
Fig.~\ref{fig:resoeffi}. The results show that
$\sigma_z^{\rm ATQsum}$ ($\epsilon_z^{\rm ATQsum}$) increases (decreases)
with increasing beam intensity. It is extremely difficult or even impossible
to determine the reaction depth for beam intensities beyond 1.6 Mcps. This
issue does not arise with the multi-layer energy-loss-difference method.
Although $\sigma_z^{\rm ATQrel;2}$ worsens slightly, increasing from about
0.18 mm at 75 kcps to about 0.23 at 1.6 Mcps, we achieved a resolution of
about 0.31 mm at 3.6 Mcps. The tracking efficiency $\epsilon_z^{\rm ATQrel;2}$
remains high at 90\% even at 3.6 Mcps.

The actual depth resolution of S$^4$AT is expected to be better than the
effective depth resolution $\sigma_z^{\rm ATQrel;2}$. As an example, we estimated
the actual depth position resolutions for S$^4$AT and the Si-CsI(Tl) telescopes
using Monte Carlo simulations taking into account the experimental
conditions. For simplicity, we considered the case with a 75-kcps proton
beam. Since the actual reaction depth position for each event is known in the
simulation, one can calculate the deviations of the simulated reconstructed depth
positions with S$^4$AT and the Si-CsI(Tl) telescopes from the actual reaction
depth position. The depth position resolutions, defined as the
root-mean-square widths of the deviation distributions, were estimated to be
about 0.14 and 0.13 mm for S$^4$AT and the Si-CsI(Tl) telescopes, respectively.


\begin{figure}[htbp] 
  \begin{centering}
    \includegraphics[scale=0.5]{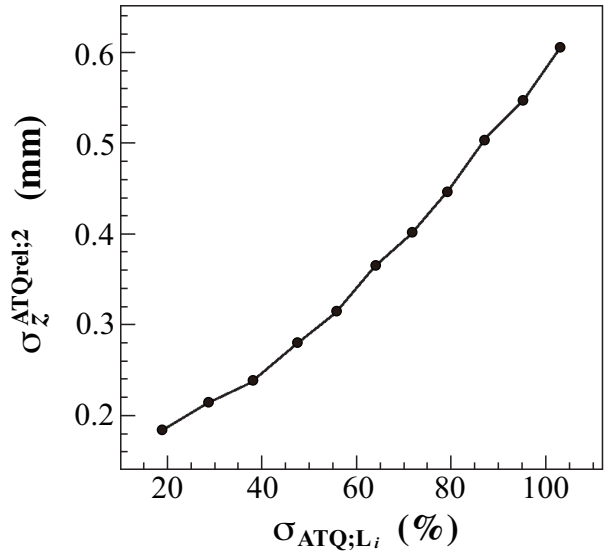}
    \caption{\label{fig:resoeffi2} Dependence of reaction depth resolution on
      intrinsic energy resolution of the plastic scintillators.}
  \end{centering}
\end{figure}
Finally, it is useful to understand the effect of the intrinsic energy
resolution of the multi-layer plastic scintillators on the depth resolution of
S$^4$AT. To investigate the dependence on the intrinsic energy resolution, we
performed Monte Carlo simulations based on the present experimental
setup. For simplicity, we assumed that all layers have the same intrinsic
energy resolutions. By artificially changing the intrinsic energy resolution,
we performed simulations to estimate the depth resolution
$\sigma_z^{\rm ATQrel;2}$. The results are plotted in Fig.~\ref{fig:resoeffi2} as
a function of energy resolution for 1-MeV energy loss of a proton in a layer of
S$^4$AT, which is denoted by $\sigma_{{\rm ATQ;L}_i}$.
As mentioned earlier, $\sigma_{{\rm ATQ;L}_i}$ was about 18\% at 75-kcps beam
intensity in the present work. This relatively poor intrinsic energy resolution
was due to light leakage into neighbouring anode strips, which resulted in a
relatively low light collection efficiency (about 60\%) in one anode strip.
Further reduction in $\sigma_{{\rm ATQ;L}_i}$ is anticipated in future
experiments by adding back the output signals from neighbouring anodes.

\section{Summary and future prospects}
\label{sec:summary}
We have constructed and tested a novel prototype solid-state active proton
target, named Stack Structure Solid organic Scintillator Active Target
(S$^4$AT), for use in nuclear spectroscopic studies with nuclear reactions in
inverse kinematics. The solid active target consists of five layers of plastic
scintillators, each with a 1-mm thickness. S$^4$AT offers the capability to 
determine the reaction depth through exploitation of the difference between the
energy losses of a charged beam particle and charged reaction products in the
scintillator material. By considering the relative energy loss between
different layers, the energy loss due to unreacted beam particles can be
eliminated. The ability to eliminate pile-up effects enables its operation at a
moderate beam intensity of up to a few Mcps. To evaluate its performance, we
have performed an elastic proton-proton measurement using a 70-MeV proton beam
at Cyclotron and Radioisotope Center (CYRIC), Tohoku University. The depth
resolution achieved was below 0.3 mm (in sigma) at 3.6-Mcps proton beam
intensity.

One of the two remaining issues is the gain shift, which will be the most
critical issue for operation at higher beam intensity. To address this issue,
we plan to introduce additional power supply boosters to the last few stages of
the MAPMT. The other issue is related to its performance with different types of
beam. The present test experiment was performed using a proton beam from
a cyclotron. It will be useful and important also to perform a test experiment
using a beam from a synchrotron accelerator, because of the difference in the
time structure of the beam, i.e. the pile-up frequency.

The present system offers the prospect of a relatively thick target while
maintaining a good energy resolution. It is worth noting that by replacing the
plastic scintillators with the deuterated ones such as the
BC436~\cite{Frenje96}, one can also construct a solid-state active deuteron
target system.

\section*{Acknowledgment}
The authors thank the CYRIC operators for the stable proton beam and the CYRIC
administrative staff for support. We also acknowledge hardware support from
T.~Furuno and J.~Zenihiro. This work was partially supported by
Hirose International Scholarship Foundation, the National Natural Science
Foundation of China under Contracts No. 11235002, No. 11375023, No. 11475014,
and No. 11575018, and the National Key R \& D program of China (2016YFA0400504).

\section*{References}

\bibliography{solid-active-target_arxiv}

\end{document}